\documentclass[twocolumn, prb, superscriptaddress]{revtex4}
\usepackage{mathrsfs}

\usepackage{amssymb}
\usepackage{graphicx}
\usepackage{dcolumn}
\usepackage{bm}
\usepackage{amsmath}

\def\OMIT#1 {{}}
\def\MEMO#1 {{}}

\makeatletter

\newcommand{\Rmnum}[1]{\expandafter\@slowromancap\romannumeral #1@}
\makeatother

\newcommand{\sjtu} {Key Laboratory of Artificial Structures and Quantum
Control, School of Physics and Astronomy, Shanghai Jiao Tong University, Shanghai 200240, People's Republic of China}

\newcommand{\WQCASQC} {Wilczek Quantum Center, School of Physics and Astronomy, Shanghai Jiao Tong University, Shanghai 200240, China}

\begin{document}

\title{Symplectic ferromagnetism and phase transitions in multi-component fermionic systems }

\author{Zi Cai}
\affiliation{\WQCASQC}
\affiliation{\sjtu}

\author{Congjun Wu}
\affiliation{Department of Physics, University of California, San Diego, CA92093}

\begin{abstract}
In this paper, we study the itinerant ferromagnetic phase in multi-component fermionic systems with symplectic  (Sp(4), or isomorphically SO(5)) symmetry.  Two different microscopic models have been considered and an effective field theory has been proposed to study the critical behavior of the nonmagnetism-magnetism  phase transition.  It has been shown that such systems exhibit  intriguing ferromagnetism and critical behavior that different from those in  spin-$\frac 12$ fermionic systems, or in high-spin systems with SU(N) symmetry. An extension of our results to higher spin systems with Sp(2N) symmetry has also been discussed.
\end{abstract}

\maketitle

\section{Introduction}

The origin of magnetism is rooted in quantum mechanics.  In transition metals like cobalt, iron and nickel, the spins of itinerant electrons are aligned to form a ferromagnetic state\cite{Vollhardt2001}. The mechanism of such an itinerant ferromagnetism can be qualitatively understood by a  simple model proposed by Stoner\cite{Stoner1933}, which predicts a spontaneous spin polarization (ferromagnetism)  in a spinful fermionic system with sufficiently large repulsive interaction,  or with a high density of states at the Fermi surface. Such a simple mean-field analysis, even though can capture some qualitative features, is not adequate to provide a quantitative description of this strongly correlated phenomenon, especially the critical behavior of the model. The nonmagnetism-magnetism phase transition in Stoner model, as  a prototypical example of quantum critical phenomena,  has attracted considerable interest in the condensed matter physics for decades\cite{Hertz1976,Millis1993}. In contrast to their classical counterpart, the quantum critical behaviors, even the static ones,  are affected by the value of dynamic critical exponent $z$ due to the inextricable connection between dynamics and statics in quantum systems. Even though a direct application of the Stoner model to  ferromagnet in solid state experiments is difficult due to the incredibly complex electron structure, the advantages of ultracold atomic systems, especially the tunablity of interactions, allows us to realize the Stoner model in such synthetic quantum systems\cite{Jo2009}.

The versatile tunability and unprecedented degree of precision of ultracold atom gases provide a perfect simulation platform to understand quantum many-body problems that have remained open, but also allow us to explore the emerging novel physics originated from
the unique features of ultracold atom systems and absent in solid state experimental systems. As an example, ultracold atom provides an opportunity to investigate the many-body effect of the atoms with total angular momentum $F$ larger than $\frac 12$, leading to $2F+1$ hyperfine states.
In solid state system, a high spin object is a composite particle composed of partially-filled shell electrons due to Hund's rule. The dominant spin exchange interaction between these high-spin objects involves only a pair of electrons, thus remain SU(2) symmetric and the quantum fluctuation is suppressed by the high spin effect(large-S limit). In ultracold atoms with high spin, on the contrary,  all the $2F + 1$ hyperfine levels could equivalently participate atom-atom exchange interactions (SU(N) symmetric), thus quantum fluctuations are enhanced by the large number of hyperfine-spin components N (large-N limit)\cite{Wu2010a},  which gives rise to a whole host of novel phenomena including unconventional
superfluidity\cite{Yip1999,Ho1999,Lecheminant2005} and exotic quantum magnetism\cite{Wu2003,Honerkamp2004,Wu2005a,Wu2006a,Qi2008,Cazalilla2009,Xu2009,Gorshkov2010,Hermele2009,Santos2010,Cai2013,Cai2013b,Wang2014}.
Motivated by rapid experimental progress\cite{DeSalvo2010,Taie2010}, ultracold fermionic gases with high spin have opened up new possibilities to explore the novel quantum many-body physics absent in solid state systems,  and have attracted a rapidly growing interests.

In this paper, we study the itinerant ferromagnetism for the ultracold fermions with spin-$\frac 32$, as well as the nonmagnetism-magnetism phase transitions associated with it. For alkaline earth atoms (e.g. $^{87}$Sr, and $^{173}$Yb), the interactions among them are independent on their hyperfine spin, which gives rise to the SU(N) symmetry of the system, and the itinerant ferromagnetism in an SU(6) fermonic system has been investigated\cite{Cazalilla2009}.  However,  for a more general case of interacting fermi gases with spin-$\frac 32$ under s-wave Feshbach resonance, it is known that the SU(4) symmetry is not generic, instead, there is a hidden SO(5) symmetry, or isomorphically, Sp(4) symmetry without fine
tuning\cite{Wu2003,Wu2006a}. This subtle difference leads to important consequences on the nature of itinerant ferromagnetism as well as the critical behavior of the phase transitions. For instance, by analogy to the spin-$\frac 12$ system, one may expect a spin-$\frac 32$ ferromagnetism with the population of one spin component larger than the other three. This kind of ferromagnetism, though possible for the SU(4) interaction, is not allowed in the ferromagnetic phases of  our system with SO(5) symmetry, as we will show in the following. Another difference lies in the effective field theory and associated phase transitions: in the SU(4) (or generally SU(N) for $N>2$) case, the phase transition is the first order due to the appearance of the cubic terms\cite{Cazalilla2009}, which, on the other hands, are absent in the effective field theory of the SO(5) systems due to the time-reversal symmetry. We have studied the critical behavior of the nonmagnetism-magnetism phase transitions in the SO(5) system.

\section{Effective field theory of symplectic ferromagnetism}
\label{sec:effective}
In this section, we propose an effective field theory for the ferromagnetism in the spin-$\frac 32$ system with SO(5) symmetry, and use it to analyze the critical behavior of the magnetism-nonmagnetism phase transition. We only focus on  spin degree of freedom, while  the charge degree of freedom of the fermions will be discussed in the subsequential section. Instead of deriving from a microscopic Hamiltonian, here we propose an effective field theory based on symmetry analysis. As a comparison, we also study a spin-$\frac 12$ system with SU(2) symmetry and spin-$\frac 32$ case with SU(4) symmetry.

Before discussing the high spin systems, we first review a ferromagnetic phase in a spin-$\frac 12$ case with SU(2) symmetry, which is characterized by a $2\times2$ matrix defined as: $\mathcal{S}(\mathbf{r})=s_x(\mathbf{r})\hat{\sigma}_x+s_y(\mathbf{r})\hat{\sigma}_y+s_z(\mathbf{r})\hat{\sigma}_z$ with $\hat{\sigma}_{x,y,z}$ being the three Pauli matrices, and the real three-dimensional (3D) vector $\vec{S}(\mathbf{r})=[s_x(\mathbf{r}),s_y(\mathbf{r}),s_z(\mathbf{r})]$ being the order parameter of the ferromagnetic phase. In general, an effective field theory of the free-energy function with SU(2) and time-reversal symmetries can be written in terms of $\mathcal{S}(\mathbf{r})$ as:
\begin{equation}
\mathcal{F}_{SU(2)}=\int d\mathbf{r}\big\{\frac{-\nabla^2+r}2\mathrm{Tr}\mathcal{S}^2+\frac{u}{4\mathcal{V}} [\mathrm{Tr}\mathcal{S}^2]^2+\cdots\big\} \label{eq:SU(2)}
\end{equation}
where $\mathcal{V}$ is the volume of the system, $r$ and $u$ are the parameters determined by the microscopic Hamiltonian and temperature. Notice that $\mathcal{S}(\mathbf{r})$ is not invariant under the time reversal transformation: $\mathcal{T}\mathcal{S}(\mathbf{r})\mathcal{T}^{-1}=-\mathcal{S}(\mathbf{r})$ where $\mathcal{T}=i\hat{\sigma}_y\mathcal{C}$ is the time reversal transformation operator, and  $\mathcal{C}$  is the operator of complex conjugation. As a consequence, the odd terms of $\mathcal{S}(\mathbf{r})$ are absent in Eq.(\ref{eq:SU(2)}). The higher order terms are neglected since they are irrelevant for the critical properties.  Notice that Eq.(\ref{eq:SU(2)}) can be rewritten in terms of $\vec{S}(\mathbf{r})$, which gives rise to a n-component real scalar field ($\varphi^4$) model with $n=3$, whose critical properties are known to be determined by the Wilson-Fisher fixed points\cite{Wilson1972}.

Now we turn to a spin-$\frac 32$ case with an SU(4) symmetry. In the fundamental representation, the 15 generators of the SU(4) group (denoted as $\hat{Q}_i$ with $i=1\sim 15$) can be expressed in terms of the ($4\times4$) Dirac matrices, which could be classified into two categories according to the time-reversal symmetry: the first classes is five SO(5) vectors which are time-reversal even: $\hat{\Gamma}^1=\hat{\sigma}^y\otimes \hat{I}$, $\hat{\Gamma}^{2-4}=\hat{\sigma}^z\otimes \hat{\sigma}^{x,y,z}$, $\hat{\Gamma}^5=\sigma^x\otimes I$. The second is ten generators of SO(5) group (or isomorphically, Sp(4) group), which are time-reversal odd and defined by $\hat{\Gamma}^{ab}=-\frac i2[\hat{\Gamma}^a,\hat{\Gamma}^b]$, where $1\leq a<b \leq 5$. It is straightforward to check that $\mathcal{T}\hat{\Gamma}^{a}\mathcal{T}^{-1}=\hat{\Gamma}^{a}$, and $\mathcal{T}\hat{\Gamma}^{ab}\mathcal{T}^{-1}=-\hat{\Gamma}^{ab}$, where  $\mathcal{T}$ is  the time reversal operator for the Dirac matrices defined as  $\mathcal{T}=\hat{\Gamma}^1\hat{\Gamma}^3\mathcal{C}$. Similar with the spin-$\frac 12$ case, the free energy function a  spin-$\frac 32$ system with SU(4) symmetry can be expressed in terms of the $4\times4$  matrix $\mathcal{Q}(\mathbf{r})=\sum_{i=1}^{15} q_i(\mathbf{r}) \hat{Q}_i$ as\cite{Cazalilla2009}:
\begin{equation}
\mathcal{F}_{SU(4)}=\int d\mathbf{r}\big\{\frac{-\nabla^2+r}2\mathrm{Tr}\mathcal{Q}^2+\eta\mathrm{Tr}\mathcal{Q}^3+\frac{u}{4\mathcal{V}} [\mathrm{Tr}\mathcal{Q}^2]^2+\cdots\big\} \label{eq:SU(4)}
\end{equation}
Notice that $F_{SU(4)}$ is fundamentally different from $F_{SU(2)}$ due to the presence of the cubic term $\mathrm{Tr}\mathcal{Q}^3$, which preserves time-reversal symmetries, thus is allowed in the free energy functional.  It is known that such a cubic term will drive the continuous phase transition in the $\varphi^4$  model to a first order one, thus there is no universal critical behavior for the phase transition in such an SU(4) model.

For a spin-$\frac 32$ system with SO(5) (or isomorphically, Sp(4) ) symmetry, the ferromagnetic phase breaks the time reversal symmetry, thus could be characterized by the matrix: $\mathcal{M}(\mathbf{r})=\sum_{[ab]}m_{ab}(\mathbf{r})\hat{\Gamma}^{ab}$, where $\sum_{[ab]}=\sum_{1\leq a<b\leq 5}$ is the summation over all the ten generators of the SO(5) group ($\hat{\Gamma}^{ab}$), which are time-reversal odd. Different from the SU(4) case, the cubic term $\mathrm{Tr}\mathcal{M}^3$ now breaks the time reversal symmetry, therefore, the minimum model preserving the SO(5) and time reversal symmetry can be expanded in terms of $\mathcal{M}$ as:
\begin{small}
\begin{equation}
\mathcal{F}_{SO(5)}=\int d\mathbf{r}\big\{\frac{-\nabla^2+r}2\mathrm{Tr}\mathcal{M}^2+\frac{v}{4\mathcal {V}}\mathrm{Tr}\mathcal
{M}^4+\frac{u}{4\mathcal{V}} [\mathrm{Tr}\mathcal{M}^2]^2+\cdots\big\} \label{eq:Sp(4)}
\end{equation}
\end{small}
using the identities:
\begin{eqnarray}
 \nonumber \mathrm{Tr}\mathcal{M}^2=\sum_{[ab]}m_{ab}^2(r),\quad
 \mathrm{Tr}\mathcal{M}^4=(\mathrm{Tr}\mathcal {M}^2)^2+\sum_{a=1}^5
 \theta_a^2(r)
 \end{eqnarray}
where $\theta_a(r)=\epsilon^{abcde}m_{bc}(r)m_{de}(r)$ ($\epsilon^{abcde}$ is antisymmetric rank-5 tensor), one can express the free energy functional  in terms of the fields $m_{ab}(\mathbf{r})$ as:
\begin{equation}
\mathcal {F}_{SO(5)}=\int
d^3\mathbf{r}\sum_{[ab]}\frac{-\nabla^2+r}2m_{ab}^2+\frac
{u'}4(\sum_{[ab]}m_{ab}^2)^2+\frac {v'}4\sum_a\theta_a^2 \label{eq:SO5}
\end{equation}
For $v'=0$, this model is an n-component $\varphi_4$ (n=10) model with an SO(10) symmetry, while the last term in Eq.(\ref{eq:SO5}) breaks the SO(10) symmetry into an SO(5) one.

To study the critical properties of the finite temperature magnetism-nonmagnetism phase transition, we apply the standard renormalization group (RG)
procedure (small $\epsilon$ expansion\cite{Zinn2002} with $\epsilon=4-d$).  After performing the functional integral over modes with $\Lambda/b<|\mathbf{k}|<\Lambda$ ($\Lambda$ is the cutoff in the momentum space, and $b>1$), we can rescale the momenta as well as the field $m_{ab}$, which leads to a model with the same form of Eq.(\ref{eq:SO5}), but  different parameters $r$, $u'$, $v'$. By taking the limit of $b\rightarrow 1^+$. one can get the differential momentum shell recursion relation to the first order in $\epsilon$ as:
\begin{eqnarray}
\nonumber \frac {dr}{d\ln b}&=&2r+\frac{K_d}{(2\pi)^d}\frac {\Lambda^d}{\Lambda^2+r}(12u'+\frac 32v')\\
 \nonumber\frac {du'}{d\ln b}&=&\epsilon u'-\frac{K_d}{(2\pi)^d}\frac {\Lambda^d}{(\Lambda^2+r)^2}(18u'^2+3u'v')\\
 \frac {dv'}{d\ln b}&=&\epsilon v'-\frac{K_d}{(2\pi)^d}\frac {\Lambda^d}{(\Lambda^d+r)^2}(2v'^2+12u'v')\label{Beta}
\end{eqnarray}
where $K_4=2\pi^2$. From Eq.(\ref{Beta}), we find there are three fixed points in the RG flow diagram (we assume $\epsilon\rightarrow 0$ and $\Lambda\gg r$): ($a$) Gaussian point: $r_a=u'_a=v'_a=0$.   The stability of the fix points can be characterized by the stability exponents: $\lambda^r_a=2$, $\lambda^{u'}_a=\lambda^{v'}_a=\epsilon$, which can be derived by performing a linear stability analysis around the fixed point.  ($b$) an SO(10) fixed point: $r=-\frac{\epsilon}{3}\Lambda^2$, $u'=\frac{4\epsilon\pi^2}{9}$, $v'=0$, with the stability exponents: $\lambda^r_b=2-\frac{2\epsilon}3$, $\lambda^{u'}_b=-\epsilon$, $\lambda^{v'}_b=\frac{\epsilon}3$; In this fixed point, the system has an enlarged SO(10) symmetry. Notice that in both ($a$) and ($b$), $\lambda^{v'}>0$, which implies the instability of these fixed points. ($c$) SO(5) point: $r=-\frac{3\epsilon}{8}\Lambda^2$,
$u'=0$, $v'=4\epsilon \pi^2$, with the stability exponents: $\lambda^r_c=2-\frac{2\epsilon}3$, $\lambda^{u'}_c=-\frac{\epsilon}2$,
$\lambda^{v'}_c=-\epsilon$. Near the SO(5) point, $\lambda^r_c>0$ and $\lambda^{u'}_c<0$  $\lambda^{v'}_c<0$, which is the character of a continuous  phase transition driven by $r$.  This fix point represents a new kind of universal class of critical phenomena.

\section{Multi-component fermionic models with SO(5) symmetry}
In this section, we consider the microscopic Hamiltonian  of two multi-component fermionic models with SO(5) symmetry. One is a ultracold atomic model, the other is from condensed matter physics.
\subsection{spin-$\frac 32$ fermionic gases with s-wave Feschbach resonance}
The first model we consider is a three-dimensional spin-$\frac 32$ fermionic gases with s-wave Feschbach resonance. Due to Paul's
exclusion principle, the spin channel wave function of two identical fermions has to be antisymmetric in the s-partial wave channel, thus
only two  channels with with total angular momentum $F=0,2$ appear in the interaction. The Hamiltonian of the system reads:
\begin{eqnarray}
\nonumber H=\int d^3\mathbf{r}\{\sum_{\alpha=\pm\frac 32,\pm\frac
12}
\psi^\dag_\alpha(\mathbf{r})(-\frac{\hbar^2}{2m}\nabla^2-\mu)\psi_\alpha(\mathbf{r})+\\
g_0 P^\dag_{00}(\mathbf{r})P_{00}(\mathbf{r})+g_2 \sum_{m=\pm2,\pm1,0} P^\dag_{2m}(\mathbf{r})P_{2m}(\mathbf{r})\}\label{pp}
\end{eqnarray}
where $\psi_\alpha(\mathbf{r})$ ($\psi_\alpha(\mathbf{r})$) annihilates(creates) a fermion with spin $\alpha$ at position $\mathbf{r}$. $\mu$ is the chemical potential. $P^\dag_{Fm_F}(\mathbf{r})=\sum_{\alpha\beta}\langle\frac 32\frac 32;Fm_F|\frac 32\frac 32\alpha\beta\rangle \psi^\dag_\alpha(\mathbf{r})\psi^\dag_\beta(\mathbf{r})$ are the channel-$F$ pair operators defined through the Clebsh-Gordan(CG)
coefficient for two indistinguishable particles with spin-$\frac 32$. The absence of the interaction channels with $F=1,3$ is important for
the general SO(5) symmetry of our Hamiltonian. $g_0, g_2 >0$  for repulsive interactions.

To analyze the magnetism of  Hamiltonian.(\ref{pp}), it is more convenient to rewrite the interaction part in the particle-hole channel, instead of the particle-particle channel in Eq.(\ref{pp}). All the possible 16 bilinear operators in the particle-hole channel can be expressed by the Dirac matrices as:
\begin{eqnarray}
\nonumber \hat{N}(\mathbf{r})&=&\Psi^\dag(\mathbf{r})\Psi(\mathbf{r}),\\
\nonumber \hat{n}_a(\mathbf{r})&=&\frac
12\Psi^\dag(\mathbf{r})\hat{\Gamma}^a\Psi(\mathbf{r}),\\
\hat{L}_{ab}(\mathbf{r})&=&-\frac 12
\Psi^\dag(\mathbf{r})\hat{\Gamma}^{ab}\Psi(\mathbf{r}).\label{SO5}
\end{eqnarray}
where $\Psi(\mathbf{r})=[\psi_{\frac 32}(\mathbf{r}),\psi_{\frac
12}(\mathbf{r}),\psi_{-\frac12}(\mathbf{r}),\psi_{-\frac32}(\mathbf{r})]^T$
is a four-component spinor operator, and the interaction part of Hamiltonian.(\ref{pp}) can be rewritten in particle-hole channels as\cite{Wu2006a}:
\begin{small}
\begin{equation}
H_{I}=\int d^3\mathbf{r} \{ f_s(N(\mathbf{r})-2)^2+f_v\sum_{a=1}^5n_{a}^2(\mathbf{r})+ f_t\sum_{[ab]} L_{ab}^2(\mathbf{r})\}\label{ph}
\end{equation}
\end{small}
where $\sum_{[ab]}$ is the summation over the index of all the ten auxiliary fields. The coefficients $f_s$, $f_v$ and $f_t$ can be written in terms of $g_0$ and $g_2$ as $f_s=(g_0+5g_2)/16$, $f_v=(g_0-3g_2)/4$ and $f_t=-(g_0+g_2)/4$. The SU(4) symmetry can be restored if $g_2=g_0$. For generic $g_0$ and $g_2$, the interaction Hamiltonian.(\ref{pp}) preserve an SO(5) symmetry without fine-tuning any parameters in Eq.(\ref{ph}).

\subsection{A two-orbital spin-$\frac 12$ fermionic Hubbard model}
In solid state physics, a multi-component fermionic system can be realized by introducing degrees of freedom other than spin, for instance, the orbital degree of freedom. To realize a four-component fermionic system, one can consider a two-orbital fermionic Hubbard model with the Hamiltonian:
\begin{eqnarray}
\nonumber H&=&\sum_{\langle ij\rangle} \sum_{\sigma a}-t(c_{i,\sigma a}^\dag c_{j,\sigma a}+h.c)+\sum_i \{-\mu n_i\\
&+&\sum_a U n_{i,\uparrow a}n_{i,\downarrow a}+V n_{i,1}n_{i,2}+J\vec{S}_{i,1}\cdot \vec{S}_{i,2}\} \label{eq:fermion}
\end{eqnarray}
where $\langle ij\rangle$ indicates a pair of adjacent sites. $\sigma=\uparrow,\downarrow$ ($a=1,2$) is the spin (orbital) index, and $c_{i,\sigma a}$ ($c_{i,\sigma a}^\dag$) is the annihilation(creation) operator of fermions at site i with spin $\sigma$ and orbital $a$.    $t$ denotes the single particle hopping amplitude, which is assumed to be independent of the spin or orbital degrees of freedom. $n_{i,\sigma a}$ is the density operator of the fermions on site i with spin $\sigma$ and orbital $a$, $n_{i,a}= n_{i,\uparrow a}+n_{i,\uparrow a}$ and $n_i= n_{i,1}+n_{i,2}$. $\mu$ is the chemical potential, and U (V) denotes the strength of the on-site intra-orbital (inter-orbital) interaction. $\vec{S}_{i,a}$ is the spin operator on orbital a, which can be written in terms of fermionic operators as: $\vec{S}_{i,a}=c^\dag_{i,\sigma a}\vec{S}_{\sigma\sigma'}c_{i,\sigma' a}$ where $\vec{S}=\frac 12 \vec\sigma$ with $\vec{\sigma}=[\sigma_x,\sigma_y,\sigma_z]$ are the Pauli matrices. $J$ represents the strength of the inter-orbital spin exchange interaction.

Such a two-orbital spin-$\frac 12$ model can be mapped to a spin-$\frac 32$ model fermionic model as: $[\psi_{{\frac 32}},\psi_{{\frac 12}},\psi_{{-\frac 12}},\psi_{{-\frac 32}}]\rightarrow [c_{\uparrow1},c_{\downarrow1},c_{\uparrow2},c_{\downarrow2}]$, where we omit the site index $i$. As a consequence, the Ham.(\ref{eq:fermion}) can be rewritten in terms of the bilinear operators in the particle-hole channel defined in Eq.(\ref{SO5}) as\cite{Wu2005}:
\begin{small}
\begin{eqnarray}
\nonumber H=\sum_{\langle ij\rangle}\sum_{\alpha}-t(\psi_{i,\alpha}^\dag\psi_{j,\alpha}+h.c)+\sum_i\{-\frac{f_0}8(\hat{N}(i)-2)^2 \\
-\mu \hat{N}(i)-\frac{f_1}8[\hat{n}^2_1(i)+\hat{n}^2_5(i)]-\frac{f_2}8[\hat{n}^2_2(i)+\hat{n}^2_3(i)+\hat{n}^2_4(i)]\} \label{eq:fermion2}
\end{eqnarray}
\end{small}
where $\hat{n}_a(i)=\Psi_i^\dag \hat{\Gamma}^a \Psi_i$ with $a=1\sim 5$, and $f_{0,1,2}$ can be written in terms of U,V,J as: $f_0=\frac 34J-U-3V$,   $f_1=\frac 34J-U+V$, $f_2=\frac 14J+U-V$. Eq.(\ref{eq:fermion2}) indicates that  different from the model of spin-$\frac 32$ fermions with s-wave Feschbach resonance, the two-orbital fermionic Hubbard model in Ham.(\ref{eq:fermion}), in general, does not possess the SO(5) symmetry, which can only be restored when $f_2$ is tuned to be the same as $f_1$, or $J=2(U-V)$. Notice that in the lattice system, the 16 bilinear operators in the particle-hole channel satisfy the Fierz identity $\sum_{[ab]}\hat{L}_{ab}^2(i)+\sum_a \hat{n}_a^2(i)+\frac 54 [\hat{N}(i)-2]^2=5$, which allows us to rewrite Eq.(\ref{eq:fermion2}) in terms of $\hat{L}_{ab}(i)$ and $\hat{N}(i)$ in the presence of SO(5) symmetry ($f_1=f_2$).

\section{Symplectic itinerant ferromagnetism and the phase transitions}

\subsection{Classical and quantum criticality}
The itinerant ferromagnetic system will experience a magnetism-nonmagnetism phase transition induced by thermal or quantum fluctuation. To study the nature of the symplectic itinerant ferromagnetism and phase transitions associated with it, we refocus on the spin-$\frac 32$ fermionic model with the interaction defined in Eq.(\ref{ph}). We are interested in  a time-reversal symmetry breaking ferromagnetic phase, whose order parameters are time reversal odd, thus can only be expressed in terms of $L_{ab}$ instead of $n_a$ and $N$. As a consequence, we will focus on the regime with $f_s,f_v>0$ and $f_t<0$, where the contribution of first two terms of th right side of Eq.(\ref{ph}) can be neglected  at the mean-field level (e.g. $\langle n_a\rangle=0$) . Using Hubbard-Stratonovich transformation, one can decouple the interaction Hamiltonian.(\ref{ph}) by  introducing ten auxiliary fields $m_{ab}(r)$ ($1\leq a<b\leq 5$), where $r=[\mathbf{r},\tau]$ and $\tau$ denotes the imaginary time satisfying $0\leq\tau<\beta$. The partition function can be written in the path integral formalism as:
\begin{small}
\begin{eqnarray}
\nonumber Z&=&\int\mathfrak{D}[m_{ab}]\mathfrak{D}[\Psi,\bar{\Psi}]e^{-\int dr\{\bar{\Psi}[\frac\partial{\partial\tau}-\frac{\hbar^2}{2m}\nabla^2-\mu]\Psi+\bar{H}_I(\Psi,m_{ab})\}},\\
\bar{H}_I&=&\sum_{[ab]}\frac12 m^2_{ab}(r)+\sqrt{\frac{-f_t}2}\bar{\Psi}(r)\Gamma^{ab}\Psi(r) m_{ab}(r)
\label{partition}
\end{eqnarray}
\end{small}
where $\int\mathfrak{D}[m_{ab}]$($\int\mathfrak{D}[\Psi]$) denotes the functional integral over the auxiliary field $m_{ab}(r)$ (Grassmann field $\Psi(r)$) and $\int dr=\int d^3\mathbf{r}\int^{\beta}_0d\tau$. $\bar{H}_I$ is the interaction Hamiltonian after the Hubbard-Stratonovich transformation. Integrating out of the fermionic degrees of freedom, one can obtain an effective partition function in terms of $m_{ab}(r)$:
\begin{equation}
Z=Z_0\int\mathfrak{D}m_{ab}(r)e^{-\frac 12\int dr \sum_{[ab]} m_{ab}^2(r)+ \mathrm{Tr}\ln(1-G^0\mathcal {M})}\label{log}
\end{equation}
where $Z_0$ ($G_0$) is the partition function (Green's function) of the noninteracting fermions. $\mathcal {M}=\sum_{[ab]}m_{ab}\Gamma^{ab}$. Expanding Eq.(\ref{log}) in a power series of $\mathcal {M}$,  one can get a general expression of free-energy function with the same form of Eq.(\ref{eq:SO5}), except that the integral is not only performed in the 3D real space, but also in the imaginary time due to the presence of quantum fluctuation.

At sufficiently high temperature, the quantum fluctuations are suppressed, thus the field $m_{ab}$ are independent of $\tau$. As a consequence, the free energy functional is the same as Eq.(\ref{eq:SO5}), based on which we can analyze the critical behavior of the finite-temperature phase transition, as we did in Sec.\ref{sec:effective}.  At zero temperature, a quantum phase transition in a spin-$\frac 12$ itinerant fermionic system  is a prototypical example of quantum critical phenomena, which has been studied by Hertz\cite{Hertz1976} and Millis\cite{Millis1993}. Here, we will generalize these results to our spin-$\frac 32$ case.  Similar with the finite temperature phase transition, the absence of the cubic terms of the SO(5) generators $\mathrm{Tr} \mathcal{M}^3$ allows us to study the quantum critical behavior of this system. It is known that at zero temperature, the quantum critical behavior is not only determined by the fluctuations in spacial dimensions, but also in the temporal one. Generally, spacial and temporal dimensions are not necessarily equivalent in the free-energy functional, and the difference is characterized by a dynamic exponent $z$, which is
determined by the way that the frequency enters the free-energy functional\cite{Sachdev1999,Vojta2007}. To get the value of $z$, we
focus on the quadratic term in free energy function:
\begin{equation}
 \mathcal{F}^{(2)}=\frac
 12\int d^3\mathbf{k}d\omega \sum_{[ab]}r(\mathbf{k},\omega)|m^{ab}(\mathbf{k},\omega)|^2
\end{equation}
the quadratic coefficient $r$ can be expressed as: $r(\mathbf{k},\omega)=1-f_t\chi(\mathbf{k},\omega)$ where $\chi$ is the vacuum polarization: $\chi(\mathbf{k},\omega)=-\int d^3\mathbf{q}d\omega' G^0(\mathbf{q},\omega')G^0(\mathbf{q}+\mathbf{k},\omega'+\omega)$. Notice that the free fermion Green's function $G^0(\mathbf{q},\omega)$  is independent of the spin of the fermionic system. As a consequence, at least in the level we consider here,  $\chi(\mathbf{k},\omega)$ should be the same with that in the spin-1/2 case, so is dynamical critical exponent $z=3$. Notice that in our case, $z+d>4$, which means that the quantum critical behavior is governed by the Gaussian point $r=0$, and the quartic and higher order terms in the free energy functional are irrelevant near the quantum critical point.

\subsection{The symplectic itinerant ferromagnetic phases and their symmetries}
The ferromagnetic phase in a spin-$\frac 32$ system is characterized by its order parameter  $\langle \mathcal{M}\rangle\neq 0$. From Eq.(\ref{eq:SO5}), one can find that  the free energy would be minimized by a non-zero $\langle\mathcal{M}\rangle$ when $r<0$. The structure of the spin-$\frac 32$ ferromagnetism is determined by the quartic terms in Eq.(\ref{eq:SO5}). To study the nature of the ferromagnetic phase in the spin-$\frac 32$ system, we first diagonalize the matrix $\mathcal{M}$, and obtain its four eigenvalues: $\pm\lambda_1$ and $\pm\lambda_2$, where $\lambda_1$ and $\lambda_2$ depend on the values of $m_{ab}(r)$. Notice that the eigenvalues of $\mathcal{M}$ can only appear in $\pm$ pairs, which is the character of the SO(5) generators. The diagonal matrix $\mathcal{D}$ can be decomposed as:
\begin{equation}
\mathcal{D}=\frac{\lambda_1+\lambda_2}2\Gamma^{15}+\frac{\lambda_1-\lambda_2}2\Gamma^{23}
\end{equation}
where  $\Gamma^{15}=\sigma^z\otimes \mathbf{I}$ and $\Gamma^{23}=\mathbf{I}\otimes \sigma^z$ are the two diagonal matrices among the SO(5) generators.

The free energy could be rewritten in terms of $\lambda_1$ and $\lambda_2$ as:
\begin{equation}
\mathcal{F}=r (\lambda_1^2+\lambda_2^2)+\frac{v}{2}(\lambda_1^4+\lambda_2^4)+u(\lambda_1^2+\lambda_2^2)^2\label{quartic}
\end{equation}
In general, the appearance of ferromagnetic phase demands $r<0$ and $u>0$. The minimization of the free energy Eq.(\ref{quartic}) depends on the
sign of $v$. For $v>0$, the free energy is minimized by $\lambda_1=\pm\lambda_2$. If we assume that the ferromagnetism is polarized  along z-axis, one can prove that $\mathcal{M}\propto \Gamma^{15}$ for $\lambda_1= \lambda_2$, and $\mathcal{M}\propto\Gamma^{23}$ for $\lambda_1=-\lambda_2$.  These two minimum correspond to two ferromagnetic phases characterized by the order parameters $M_1\propto \langle L_{15}(\mathbf{r})\rangle$ or $\langle
L_{23}(\mathbf{r}\rangle)$ respectively. Notice that these two ferromagnetic phases can be connected with each other by performing a SO(5) rotation in the spin space, which indicates that they are belong to the same kind of ferromagnetism. This is analogue to the spin-$\frac 12$ system with SO(3) symmetry, where the ferromagnets polarized along the x- or z-directions are belong to the same phase.

\begin{figure}[htb]
\includegraphics[width=0.9\linewidth]
{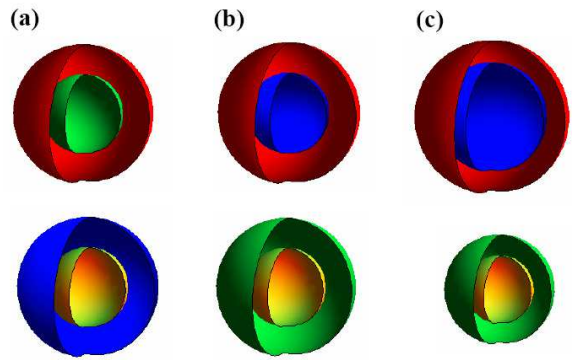} \caption{Fermi surface of fermions with different
components ($\frac 32$ red, $\frac 12$ blue, $-\frac 12$ green,
$-\frac 32$ yellow) in the itinerant ferromagnetism of spin-3/2
fermions.  (a) $\langle L_{15}\rangle\neq 0$, $\langle
L_{23}\rangle=\langle n_4\rangle=0$ ($\delta n_{3/2}=\delta
n_{1/2}=-\delta n_{-1/2}=-\delta n_{-3/2}$). (b) $\langle
L_{23}\rangle\neq 0$, $\langle L_{15}\rangle=\langle n_4\rangle=0$
($\delta n_{3/2}=\delta n_{-1/2}=-\delta n_{1/2}=-\delta
n_{-3/2}$).(c) $\langle L_{23}\rangle=\langle L_{15}\rangle\neq 0$,
$\langle n_4\rangle=0$ ($\delta n_{3/2}=-\delta n_{-3/2}$, $\delta
n_{-1/2}=\delta n_{1/2}=0$). (a) and (b) can be connected by a SO(5)
rotation thus  represent the same kind of IFM, which is different
from (c).}\label{Bloch}
\end{figure}

Besides the  spontaneous time-reversal and SO(5) symmetries breaking, this kind of itinerant ferromagnetic phases are characterized by features that the four species of the fermions can be divided into two groups, and the two species within each group are equally populated, while the population of one group is larger than the other, as shown in Fig.\ref{Bloch}.(a) and (b).  It seems plausible that
the residue symmetry in this kind of ferromagnetic phase is SU(2)$\otimes$SU(2), corresponding to the rotation invariance within the subspaces of each group. However, the overall SO(5) symmetry in our system demands that the rotations within each
subspace are not independent with the other, instead, they are connected by a U(1) phase. As a consequence,  the residue symmetry is
SU(2)$\otimes$U(1), and the manifold of the order parameter of this kind of  itinerant ferromagnetic phases is SO(5)/[SU(2)$\otimes$U(1)].

In the case of $v<0$, the free energy is minimized by $\lambda_1=0$ or $\lambda_2=0$, which corresponds to the ferromagnetic phase characterized by the non-zero order  parameter $M_2=\langle L_{15}(\mathbf{r})\rangle\pm \langle L_{23}(\mathbf{r})\rangle$. This kind of ferromagnetic phase is different from those studied previously, since  its order parameter $M_2$ can not be transformed to $M_1$ by a SO(5) rotation.  Physically, such an itinerant ferromagnetic phase are characterized by the feature that the population of the fermions in one of the four species is larger, and another one is smaller than those of the noninteracting case, while the remaining two are unchanged, as shown in Fig.\ref{Bloch} (c).

For such a four-specie fermionic system, one may expect a plausible ferromagnetic phase with the population of one species is larger than those of the other three, which are equally populated.  This kind of IFM, though possible for the SU(4) interaction, does not exists for our system with SO(5) symmetry. Without losing generality, we assume that the population of the fermions with $S_z=+\frac 32$ is larger than the other three, this itinerant ferromagnetic phase is characterized by a nonzero order parameter  $M_3=\langle L_{15}(\mathbf{r})+L_{23}(\mathbf{r})+n_4(\mathbf{r})$. $\langle n_4(\mathbf{r})\rangle=\frac 12\langle\Psi^\dag(r)\Gamma^4\Psi(r)\rangle$ ($\Gamma^4=\sigma^z\otimes \sigma^z$) denotes a spin-quadratic order\cite{Tu2006}, and does not break the time-reversal symmetry. However, within the parameter region we studied $f_v>0$ $f_t<0$, there is no spin-quadratic order $\langle n_4\rangle=0$.  Mathematically, this is because the eigenvalues of $\mathcal{M}$ can only appears in $\pm$ pairs, which
is the character of SO(5) generators.

\subsection{Higher-spin fermions with Sp(2n) symmetry}

Our discussion about the itinerant ferromagnetic phase in spin-$\frac 32$ fermions with SO(5) (or isomorphically, Sp(4)) symmetry can be generalized to higher-spin fermions with Sp(2n) symmetry. For ultracold fermion system with spin $s=n-\frac12$, a symplectic (or Sp(2n)) symmetry can be realized if one tunes the interaction parameters of all the spin multiple channels to be the same ($U_2=U_4=\cdots=U_{2n-2}$)\cite{Wu2005}. To
decouple the Sp(2n) symmetric interaction,  we introduce a $2n\times 2n$ matrix $\mathcal {M}'$, which is the analogue of $\mathcal {M}$ defined above in the Sp(4) case. Notice that similar with spin-$\frac 32$ case, the eigenvalues of $\mathcal {M}'$ can only appear as $\pm$ pairs
($\pm\lambda_1,\pm\lambda_2,\cdots,\pm\lambda_n$), and the free energy for this $Sp(2n)$ case as:
\begin{equation}
\mathcal
{F}^{Sp(2n)}=r\sum_{i=1}^n\lambda_i^2+\frac{v}2\sum_{i=1}^n\lambda_i^4+u(\sum_{i=1}^n\lambda_i^2)^2
\end{equation}
When $r<0$,  there are two different kinds of itinerant ferromagnetic phases: (a)$v>0$, the $2n$ species fermions can be classified into two groups, each of which contains n species of fermions with the same population, while one group has larger population than the other; (b)$v<0$, among the 2n species, only the population of one pair of species is changed, while the others are the same with the non-interacting case.  The phase transition can be analyzed in the similar way with the Sp(4) case.

\section{Conclusion and outlook}
In a summary, we study the itinerant ferromagnetic phases in a multi-component fermionic systems with symplectic symmetry, which gives rise to novel ferromagnetism and critical behavior that absent in spin-$\frac 12$ fermionic systems or in systems with higher spin but SU(N) symmetry. Further development includes a generalization of our results to other magnetic orders, for instance, the anti-ferromagnetic order\cite{Qi2008,Xu2009}, or to the nonmagnetic spin orders like the spin-quadratic order\cite{Tu2006} which preserve the time reversal symmetry.

\section{Acknowledgments}
ZC is supported in part by the National Key Research and Development Program of China (Grant No. 2016YFA0302001), NSFC of  China (Grant No. 11674221, No.11574200), the Project of Thousand Youth Talents, the Program Professor of Special Appointment (Eastern Scholar) at Shanghai Institutions of Higher Learning and the Shanghai Rising-Star program. 	


\end{document}